# Micro-Structuring, Ablation and Defect Generation in Graphene with Femtosecond Pulses

Andres Vasquez[1], Panagis Samolis[1], Junjie Zeng[1], Michelle Y Sander[1,2,*]

[1]*Department of Electrical and Computer Engineering and BU Photonics Center, Boston University, Boston, MA, USA*
[2]*Division of Materials Science and Engineering, Boston University, Brookline, MA, USA*
*Corresponding author: msander@bu.edu

Femtosecond micromachining offers a contact-free and mask-less technique for material patterning. With ultrafast laser irradiation, permanent modifications to the properties of single layer graphene through material ablation or defect introduction can be induced. Multiple femtosecond pulse interactions with a single layer graphene are studied and a low laser ablation threshold ~9.2 mJ/cm$^2$ is reported for a 15 second illumination time. Clean ablated structures are generated in such a multi pulse irradiation configuration at low pulse energies as an attractive alternative to ablation with single femtosecond, high energy pulses. For a fully ablated graphene hole, a radially symmetric region extending around 2 μm from the ablated edge is characterized by strong defect generation. Average distances between point-defects down to ~58 nm are derived and Raman spectroscopy implies that overall there is a strong resemblance to amorphous structures. For fluence values around 75% of the ablation threshold, modification with defect generation down to ~48 nm average defects lengths is reported, while the underlying graphene structure is maintained. Thus, depending on the laser parameter choice, the same laser configuration can be used to ablate graphene or to primarily introduce defect states. The presented findings offer interesting insights into femtosecond induced structural modifications of graphene that can lead to improved precision ablation and patterning of single-layer materials at the micro- and nano-scale. Further, this can be attractive for graphene or carbon-based device fabrication as well as sensor and transistor applications, where regions of varying carrier concentrations and different electrical, optical or physical properties are desired.

## I. INTRODUCTION

Graphene combines attractive features with high mechanical strength and good thermal and electrical conductivity. These inherent properties make graphene extremely attractive for applications with tunable electrical conductivity (e.g. in small-scale transistors and flexible electronic devices), optical modulation [1], sensors or microscale devices with tunable mechanical properties and membrane filtration. Recently, structured or patterned graphene at the micro- and nanoscale in form of ribbons, disks and other forms has particularly raised interest for applications in plasmonics or THz antennas [2]. Further, graphene kirigami [3], where three-dimensional structures are formed based on cutting and folding of graphene, has sparked widespread interest. Such three-dimensional building blocks with springs of variable mechanical strength or complex electrostatic fields can fuel the design of novel sensors, chemical and biological filtration devices, nano-catalysts or even probes for neural interactions. However, to realize the full enabling potential of graphene for all these applications, well-defined and high precision patterning of graphene is needed.

Conventional patterning methods [4] rely on electron-beam lithography, mechanical cleaving, chemical etching or focused ion beam milling to modify substrates like graphene. However, exposure to lithography process steps and the associated chemicals bears the risk of contamination and unintentional doping of the graphene, which can induce changes in its electrical and optical properties. In addition, dangling bonds instead of atomically smooth edges can be formed, which in turn can induce inhomogeneous electrostatic or magnetic potentials. This can lead to chemical activity or self-passivating reconstructive efforts. Mask-less and contact-free processing is further attractive to enhance flexibility of patterns and dynamic programming of micro- to nano-scale features. Additional wafer size restrictions for photo-lithography or electron-beam lithography can limit the processing of chip-scale or individual sensor-based systems, whereas such restrictions do not apply to laser irradiation.

Femtosecond (fs) micromachining for advanced materials processing [5–7] has a strong history of enabling high precision cuts with minimal thermal exposure of the surrounding environment in a process that avoids chemical contamination. Laser micromachining can be scaled up to arbitrary shapes, it can be used for high throughput processing and it can be applied to free-standing graphene [8], curved shapes (e.g. carbon nanotubes) or flexible films [9]. Thus, femtosecond ablation offers as significant potential for enabling high precision patterning of graphene for micro- or nanoscale devices.

Recently, selective work on experimental optimization of the ablation process for graphene has emerged [10–14], where the ablation process dependency on pulse intensity fluences has been studied for different laser systems. Most graphene micromachining research so far in the near-infrared wavelength regime around 800 nm has mostly been conducted with highly energetic femtosecond pulses with a minimum of a few µJ energy at kHz repetition rates or lower [10,11,15], requiring powerful and cost-intensive laser configurations. Singel 100 fs pulse interaction with graphene (amplified Ti:sapphire laser with a 1 kHz repetition rate) [16], resulted in petal-like graphene folding around ablated spots for fluence values between $108 - 326$ mJ/cm$^2$ with an ablation threshold at 98 mJ/cm$^2$. Utilizing one-dimensional phase-shifting plates and two-dimensional vortex plates into a 100 fs Ti:sapphire laser at a repetition rate of 76 MHz, graphene nanopatterning with nanoribbons down to 20 nm was demonstrated [8]. Using Bessel beams, the alignment constraints for the sample placement for nanostructuring of graphene were significantly reduced [13]. The single-shot exposure threshold of 156 mJ/cm$^2$ agreed well with reported values from other groups, while for multi-pulse irradiation (for less than 100 pulses), the ablation threshold was significantly reduced [13]. Studies relying on multi-pulse exposure with short 50 fs laser pulses, where optically induced pear-shaped damage to graphene was reported for pulse fluences as low as 14 mJ/cm$^2$, corresponding to maximum intensity values of 16 kW/cm$^2$ [17] for illumination times up to 5 seconds. Multi-pulse ablation at twice the former fs pulse duration (1000 pulses with 100 fs duration at 1 kHz repetition rate) with a peak fluence ~1 J/cm$^2$ for a calculated beam spot radius around 2 µm, lead to the generation structural modifications within graphene [18]. Increasing the pulse duration, picosecond functionalization and ablation of graphene at a wavelength of 515 nm for pulse energies between $2.55$ µJ – $2.85$ µJ resulted in clearly cut lines between $2 - 7$ µm width at a scanning speed of 0.15 m/s [19]. Ablation at other wavelengths, e.g. at 248 nm with 20 ns pulses [20], at 1030 nm with 280 fs pulses resulted in a single-shot damage threshold of 139 mJ/cm$^2$ [12] or at 1064 nm with 400 fs multi-pulse interaction on a silicon wafer to values of 76 mJ/cm$^2$ [21]. Studies on picosecond pulse ablation at different wavelengths were conducted as well in comparison to femtosecond pulses [22]. Although the impact of various laser pulse parameters has been studied, single pulse modification can be very sensitive to perturbations, energy fluctuations and alignment. Due to the ablation thresholds on the order of ~100 mJ/cm$^2$, most of the single and even multi-pulse irradiation configurations need higher energy pulses which require more complex and amplified laser configuration systems.

In this paper, we aim to study the interaction of multi-pulse femtosecond laser pulses with lower pulse energies starting at 15 nJ. We demonstrate that instead of single pulse ablation, interaction with multiple pulses with lower fluence values around 23.6 mJ/cm$^2$ to 43.3 mJ/cm$^2$ at 80 MHz repetition rate can result in similarly clean graphene patterning at much reduced peak intensities. The laser ablation threshold around 9.2 mJ/cm$^2$ is the lowest value reported to date, to the best of our knowledge. It can be conducted with a standard Ti:sapphire laser and the multi-pulse exposure adds some robustness to the processing due to the longer material interaction time. Raman spectroscopy enabled detailed insight into the defect structure close to the ablation edge. We show that for micromachining with fluence values below the ablation threshold, defects in graphene are generated. The discussed findings can lead to a more precise tailoring of the ablation characteristics through temporal shaping, modulation of the pulse train envelope and enhanced control of the deposited energy for higher precision patterning and chemical modification of graphene.

## II. FEMTOSECOND MICROMACHINING SET-UP

A femtosecond Ti:sapphire laser (MaiTai Spectra Physics) with a repetition rate of 80 MHz and ~100 fs pulse duration at a wavelength of 819 nm is used for the graphene ablation. A schematic of the optical setup is illustrated in Fig. 1(a). In a free-space coupling configuration, the laser beam is focused with a plano-convex lens with a focal length of 11.6 mm onto the sample. Commercially purchased chemical vapor deposition (CVD) graphene was transferred to a silicon wafer with a thin silicon oxide layer on top, attached to a microscope slide. Spatial patterning of the graphene sample is achieved via two DC servo motor controllers. Through a connected computer interface, random material patterns with high flexiblity can be programmed.

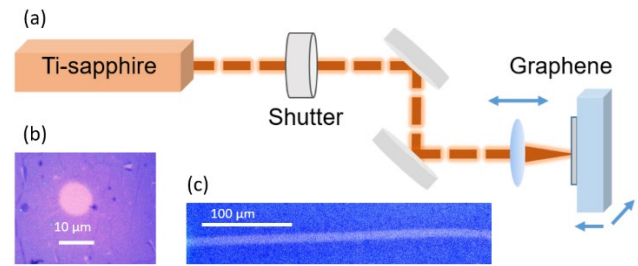

FIG. 1. (a) Micromachining set-up including an 80 MHz Ti:sapphire laser with 100 fs pulse duration focused onto the graphene sample. (b) Optical microscope image of a single ablated graphene dot shows a clean ablated area. (c) Optical image of ablated line on graphene layer.

For the studies conducted below, each imaging spot size is ablated for 15 seconds at varying power levels as indicated. While initial experiments were conducted with a movable beam block, an automated shutter for more reproducible data acquisition was incorporated. The power level was varied between 1.2 W to 2.2 W, corresponding to pulse energies between 15 nJ – 27.5 nJ per pulse. For the sample alignment, the sample was translated 0.003 inches or 76 μm from the focus position where the Gaussian beam parameter has expanded to a spot size diameter of ~12.7 μm. At the exact focus, strong visible light generation and oxidation was noticeable otherwise. By keeping the relative position of the sample compared to the focusing lens fixed at 76 μm, the incident fluence on the graphene sample was changed by altering the incident power.

Optical characterization is performed in a Nikon Eclipse LV150 microscope. Fig. 1(b) shows the microscope image of a single ablated graphene hole with 100x magnification. A symmetrically ablated spot is visible with clearly defined and well-cut edges. By translating the stage at a speed of 1 μm/s, a clean cut with over 370 μm length is written into the graphene layer, cf. Fig 1(c) with an optical microscope image with 20x magnification. The width of the ablated line ~8.4 μm remains fairly constant over its length for an incident fluence value of 32 mJ/cm$^2$.

## III. CHARACTERIZATION OF MICRO-STRUCTURED GRAPHENE AND DISCUSSION

### A. Ablation Characterization

To characterize the performance of the micromachining set-up, optical microscopy images were analyzed to determine the ablated spot size. As the extent of chemical modification and the generation of defect states is critical for high quality precision cutting and material modification, detailed Raman spectroscopy studies were conducted. The ablated spot diameter was determined by fitting a circle to the inner boundary of the ablated spot in the microscope images. Three or more ablated holes under the same laser irradiation conditions were routinely compared to ensure the reproducibility of the micromachining results.

Fig. 2(a) shows a sequence of ablated graphene dots that were illuminated with different laser power values between 1.2 W and 2.2 W, with the power increasing in 0.2 W increments from left to right. The corresponding measured ablated dot diameter $D$ is plotted in Fig. 2(b). We observe that even though the calculated theoretical Gaussian beam spot size $2 \cdot w_0$ is the same for all measurements ~12.7 μm, the measured ablated dot diameter increases for higher incident fluence values. For the studied peak fluence values $F_0$ from 23.6 mJ/cm$^2$ to 43.3 mJ/cm$^2$, the ablated dot diameter $D$ varies from 5.3 μm to 11.1 μm, respectively. This can be explained by the fact that only the fraction of the Gaussian beam above a certain fluence value threshold leads to ablation at the microscale of graphene. Based on the comparison of the Gaussian beam diameter $2 \cdot w_0$ with the actual ablated hole diameter $D$, a threshold ablation fluence $F_{th}$ value can be derived [16]:

$$D^2 = 2w_0^2 \ln(F_0 / F_{th}) \quad \text{with} \quad F_0 = \frac{P_{ave} f_{rep}}{\pi w_0^2 / 2} \quad (1)$$

For each data point, the ablation threshold fluence is calculated and plotted in Fig. 2(b). While for lower power values, only a smaller fraction of the illuminated spot area contributes to the ablated area, this fraction increases steadily towards a ratio ~0.87. Correspondingly, the ablation threshold is higher for lower pulse energy illumination, where the $F_{th}$ reaches up to values of 16.7 mJ/cm$^2$. Overall, the values converge towards a minimum threshold fluence of 9.5 mJ/cm$^2$ for higher pulse energies.

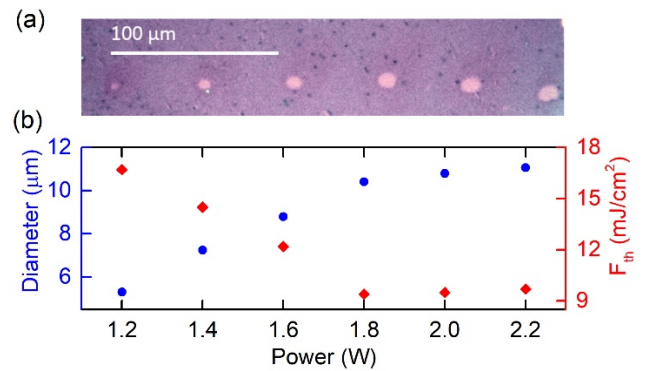

FIG. 2. (a) Optical microscope image of ablated dots in a single layer graphene for increasing power values from 1.2 W to 2.2 W. (b) Corresponding ablated dot diameter for fluence values between 23.6 mJ/cm$^2$ and 43.3 mJ/cm$^2$, respectively, which leads to a minimum ablation threshold fluence for higher pulse energies of 9.5 mJ/cm$^2$.

Another study was conducted in which the incident fluence was changed by keeping the incident power constant at 1.4 W, but altering the relative offset of the sample with respect to the previous sample focus in increments of 0.001 inches or 25.4 μm. By translating the sample by an additional 127 μm, the corresponding Gaussian beam diameter expanded from 12.7 μm to 16.6 μm. In Fig. 3(a), the measured modified graphene dot diameter is plotted against the required threshold laser fluence to induce modifications. The diameter is extracted based on different optical contrast in the microscope image, as presented in Fig. 3(b). Overall, the diameter varied between 3.7 μm to 9.2 μm with a standard deviation < ± 0.15 μm. In Fig. 3(b), the modified graphene pattern grid is shown, where each column featuring the same laser setting and diameter values are averaged based on three ablated dots. Within each row from the right to the left an additional incremental offset of 25.4 µm from the sample focus position is added, resulting in decreasing diameter graphene dots. The sample is translated by a maximum offset of 152 µm for the left most column in Fig. 3(a). While the right three columns show clearly defined ablated holes, the columns towards the left do not feature as well defined edges since the graphene is functionalized instead of fully ablated. For the ablated dots illuminated with higher fluence values, a threshold fluence between 9.2 mJ/cm$^2$ and 9.3 mJ/cm$^2$ for ablation is determined. This value agrees well with the findings from Fig. 2. However, for lower fluence values the graphene only gets modified and not fully ablated any more. Decreasing the fluence below 7.9 mJ/cm$^2$ lead to micro-modifications of graphene, however, the associated area could not be resolved with high enough contrast under the microscope any more. Thus, the left most column in Fig. 3(a) was not included in the analysis in Fig. 3(b) since the determination of a diameter based on the microscope image was ambiguous. Raman spectroscopy studies, as shown in Figs. 4 and 5 offered additional insights and confirmed that for threshold fluence values around 8 mJ/cm$^2$, the graphene is functionalized but is not fully ablated any more.

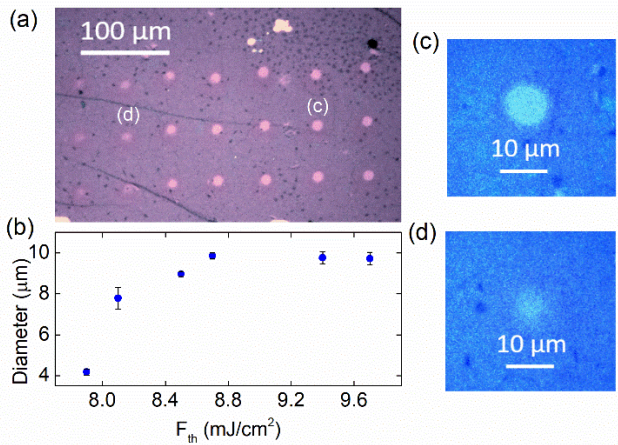

FIG. 3. (a) Optical image of ablated graphene hole grid where each column corresponds to a larger offset in 0.001 inch steps, starting from the right and increasing towards the left. (b) Evolution of measured diameter of modified graphene plotted against the required ablation threshold, corresponding to an increasing offset from the focusing lens from sample at an average power of 1.4 W. (c) Optical image of fully ablated dot with a corresponding ablation threshold of $F_{th}$ = 9.2 mJ/cm$^2$. (d) Optical image of modified graphene area with an incident fluence $F_0$ = 8.4 mJ/cm$^2$.

## B. Raman Spectroscopy

Detailed Raman spectroscopy on graphene [23,24] was conducted to characterize the quality of the underlying graphene structure and any created defect states. This was performed with a Renishaw Raman Spectroscopy instrument at a 532 nm laser wavelength (2.33 eV energy) with a power of 28.2 µW (grating 2400 l/mm and wavenumber resolution of 0.86 cm$^{-1}$). We report a detailed evolution of the Raman spectra based on radial symmetry around the graphene dots for low energy fully ablated and modified graphene spots. A fine stepping size of 0.15 µm steps was chosen for the Raman measurements. In Fig. 4(a), Raman spectra from the fully ablated graphene region (arbitrary origin) to the pristine graphene are displayed for the characteristic G, 2D (or G') and D bands. Based on radial symmetry, the evolution for just one side of the ablated hole is presented. For spatial positions below 1 µm, essentially a noise spectrum was collected, indicating that the graphene was fully ablated and removed.

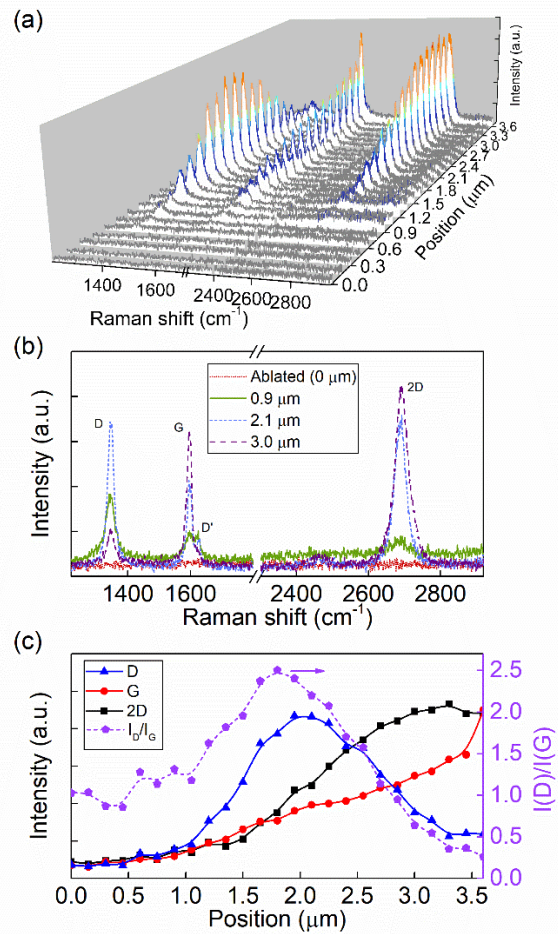

FIG. 4. Raman spectroscopy evolution of an ablated graphene hole exposed to a laser fluence of 27.5 mJ/cm$^2$. For a spatial radially symmetric offset ~1.2 µm, the D' peak signature becomes visible, indicating the generation of defects close to the hole edge. (b) Selected Raman spectra for different positions show the growth and subsequent decrease of the defect-related D band. (c) The peak intensity of each representative band for graphene is plotted together with the ratio of $I(D)/I(G)$ which characterizes defect states.

Around a position of ~1.2 μm, the D peak centered around 1345 – 1350 cm$^{-1}$ starts emerging. This Raman peak characterizes the amount of disorder close to the ablation border, indicating that the sp$^2$ graphene lattice symmetry has been broken at the ablated edge, indicative of defects [10]. The D line increases in strength up to a spatial position of ~2 μm offset from the ablated ridge. Further away from the ablated border, the D peak intensity decreases again since the edge effects are reduced, compare Figs. 4(b) and 4(c). For spatial positions beyond 3 μm, the D peak is minimized or fully disappears, so that a characteristic spectral signature of pristine graphene is measured. At the same time, the strength of the 2D peak increases until it plateaus around 3 μm. Thus, the Raman spectrum with a strong signature of the G and the 2D (or G') peaks resembles that of pristine graphene, indicating that the graphene monolayer properties have been fully preserved. The graphitic G band around 1580 – 1600 cm$^{-1}$ originates from the in-plane bond stretching of pairs of C atoms with sp$^2$ hybridization and corresponds to doubly degenerate (in-plane transverse optical, iTO, and longitudinal optical, LO) phonon modes at the Brillouin zone center [23]. The 2D band ~2700 cm$^{-1}$ is induced by second order zone-boundary phonons [25].

Based on this Raman characterization a unique transition region can be defined that extends over ~2 μm from the ablated hole edge unto the graphene layer, where the graphene has been optically modified and is characterized by increased defect states. The disappearance of the G peak until a few microns away from the hole edge, indicates that close to the ablated hole, amorphous structures can be found. In Fig. 4(b), the Raman spectrum for four selective spatial positions is presented. The appearance of the D' peak, which represents a weak disorder induced feature around 1620 cm$^{-1}$, is clearly visible. Its strength is slowly reduced up to a spatial position of ~3 μm. While the growth of the G band and the emergence of a D' peak can also be attributed to graphene oxidation [26], no significant broadening of the G band is observed here. Thus, as the shape and FWHM of the bands do not vary significantly, which would be expected from a potential graphene oxidation process, it is assumed that the generated defects are combined with potential functionalization by epoxy, hydroxyl or carboxyl groups [19]. Evaluating the intensity ratio of the peak of the D and the G bands $I(D)/I(G)$ with a maximum value close to 2 with for a spatial position ~1.8 μm indicates that there is a small range of amorphous carbon that has been generated due to optical and thermal functionalization. Consequently, these effects can be attributed to photo-induced defect generation.

From the ratio of the intensity of the D band to the G band, the average distance between point defects $L_D$ in nm can be calculated [27]:

$$L_D^2 = \frac{4.3 \cdot 10^3}{E_L^4} \cdot \left[\frac{I(D)}{I(G)}\right]^{-1} \qquad (2)$$

For a Raman excitation energy of 2.33 eV and with $I(D)/I(G)$ varying between 1 to 2.5 and then being reduced to 0.26, this indicates that $L_D$ corresponds to approximately 145 nm and gets reduced down to ~58 nm before it increases again to a fairly wide separation of 561 nm. Based on this result, the defect density $n_D$ in cm$^{-2}$ can be derived in the following way [27]:

$$n_D = \frac{1}{\pi L_D^2} = 7.3 \cdot 10^9 \cdot E_L^4 \cdot \frac{I(D)}{I(G)} \qquad (3)$$

This translates to defect densities that increase from 2.15·10$^{11}$ cm$^{-2}$ to 5.38·10$^{11}$ cm$^{-2}$ and 5.59·10$^{10}$ cm$^{-2}$ for the fairly pristine graphene. In this context, only Raman active defects are captured with this approximation for low doping values, as assumed in this case. Thus, edge contributions are represented in this value, while perfect zigzag edges, charged impurities, strain effect or intercalants do not generate any D signature, as discussed in [27,28].

In addition, a shift to lower wavenumbers for the G band up to around 6 – 10 cm$^{-1}$ from ~1597 cm$^{-1}$ for the pristine graphene down to ~1588 cm$^{-1}$ for the ablated graphene is observed. These can be related to enhanced bond disorder and amorphous structures [29]. Similarly, the 2D peak of graphene around 2694 cm$^{-1}$ is down-shifted towards 2675 cm$^{-1}$ – 2680 cm$^{-1}$ over a 2 μm spatial width. Such center wavenumber shifts can be related to enhanced electron availability and changes in doping [27]. These findings agree well with the reports in [19], where similar shifts of 5 cm$^{-2}$ in the G band were recorded and linked to doping changes [30]. For a more detailed understanding of how the measured defect states are related to absolute doping values, in-depth studies are ongoing to be combined with DFT calculations.

At the same time, for fluence levels of 8.4 mJ/cm$^2$, corresponding to a laser power of 1.4 W with an additional 127 μm offset (compare Fig. 3(d)), the threshold fluence in the center of the beam corresponds to 7.9 mJ/cm$^2$. For such a fluence value of 75% of the ablation threshold fluence, the graphene gets modified. In Fig. 5(a), the center part of the illuminated laser area has been functionalized but not ablated. A strong growth of the D band indicates functionalization of the graphene through the generation of additional defects. At the same time, the graphene lattice structure is maintained, as indicated by the remaining G peak. Thus, due to the lower fluences, no amorphization process is initiated. However, the D' peak starts increasing significantly, reaching almost similar strength as the G and 2D peak where the highest peak intensity of the Gaussian beam interacted with the sample. The detailed spectra for different spatial positions are shown in Fig. 5(b). Most characteristic is the development of the peak of the different spectral bands, cf. Fig. 5(c). A plateau region extending over a width of ~4 μm is presented, whose dimension agrees well with the extracted modified graphene diameter of 3.7 μm from the optical microscope in Fig. 3. The ratio of $I(D)/I(G)$ amounts to values between 2.5 and 3 in the functionalized area before it is reduced to values below 0.5. Thus, average distances between defects are estimated to vary between 58 nm to 48 nm, which corresponds to defect densities of $5.37 \cdot 10^{11}$ cm$^{-2}$ and $6.45 \cdot 10^{11}$ cm$^{-2}$. These are slightly higher than the defect densities generated in the transition region in Fig. 4.

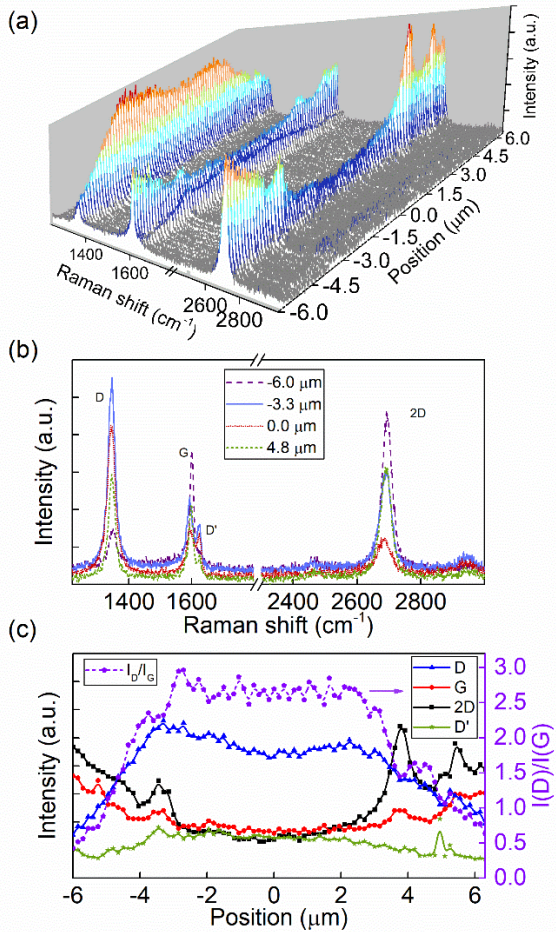

FIG. 5. (a) Raman spectroscopy on graphene exposed to a laser fluence of 8.4 mJ/cm$^2$ that resulted in a modified graphene section but not a fully ablated hole. (b) Raman spectra for different spatial positions. The D' peak is particularly strong and comparable to the D and G peaks in the modified region. The presence of a remaining G band indicates that the underlying graphene structure is maintained in this regime. (c) The peak intensity of the representative Raman bands are shown. The modified graphene region is characterized by a fairly flat plateau region extending over ~ 4 μm where the intensity ratio $I(D)/I(G)$ varies between values of 2.5 and 3.

For the modified graphene with defect states, the wavenumber shifts of the center wavelength are pronounced as well, featuring almost similar shifts as noticed before: The G band around ~1600 cm$^{-1}$ shifts down to ~1592 cm$^{-1}$ while the 2D band shifts from 2695 cm$^{-1}$ at the edges of the graphene down to 2680 cm$^{-1}$ over a region of ~5 μm with a minimum at the center of 2675 cm$^{-1}$. Thus, modification of the graphene properties without ablation was successfully induced and shown to be fairly sensitive to power variations around the threshold ablation fluence value.

## IV. CONCLUSION

In this paper, we analyzed the femtosecond laser interaction with single layer graphene to gain insights into the ablation and defect dynamics around the ablation threshold and below. Using a high repetition rate laser source with moderate pulse energies, low threshold fluences for ablation around 9.2 mJ/cm$^2$ were demonstrated, which marks the lowest values to date. Careful analysis of the ablated hole revealed clean edges extending over a zone around 2 μm width that is characterized by amorphous structures with defect densities around $5.59 \cdot 10^{10}$ cm$^{-2}$. For fluence values of 75 % of the threshold ablation fluence, defect states are induced while the underlying graphene structure is maintained. Through changing the offset of the sample position from the focusing lens, the material interaction can be switched from ablation to surface modification. The insights gathered underline that micromachining with lower pulse energies can result in well-defined patterns but can also lead to the intentional generation of defects in a controlled fashion. Enhanced spatial resolution can be obtained for micromachining of nano-ribbons and other fine features through incorporation of a tighter focusing lens. Additional studies how the defect generation relates to bandgap changes and overall carrier charge concentration are ongoing and promise to reveal interesting insights into changing optical and electrical properties of graphene. The contact-free micromachining approach without any risk for contamination and high precision patterning thus forges an attractive pathway towards 3D kirigami and other graphene applications. Similarly, it enables micro-modification of graphene and its structural properties in a flexible and versatile manner for a wide range of applications, in sensor, flexible electronics or transistor devics.

**ACKNOWLEDGEMENT**

We acknowledge Sahar Sharifzadeh, Mohammad Alaghemandi and Xi Ling for insightful scientific discussions. We thank Anna Swan for the preparation of the pristine graphene sample and Scott Bunch and his group for advice on Raman measurements. Ahmet Akosman, Adam Sapp and Peter Erf contributed to the experimental setup configuration. Funding from the Boston University College of Engineering Dean's Catalyst Award 2017 and UROP funding for A. Vasquez is acknowledged.